\documentclass[a4paper]{jpconf}
\usepackage{graphicx}
\usepackage{mathrsfs}
\usepackage{wrapfig}
\usepackage{cite}

\usepackage[usenames]{color}

\usepackage{ulem}

\begin{document}
\title{DMFT Study for Valence Fluctuations in the Extended Periodic 
Anderson Model
}

\author{Ryu Shinzaki, Joji Nasu, and Akihisa Koga}

\address{Department of Physics, Tokyo Institute of Technology, 
Tokyo 152-8551, Japan}

\ead{ryushin@stat.phys.titech.ac.jp}

\begin{abstract}
We study valence fluctuations at finite temperatures
in the extended periodic Anderson model, 
where the Coulomb interaction between conduction and localized $f$-electrons
is taken into account, 
using dynamical mean-field theory combined with 
the continuous-time quantum Monte Carlo (CT-QMC) method. 
The valence transition with the hysteresis is clearly found, 
indicating the first-order phase transition 
between the Kondo and mixed-valence states.
We demonstrate that spin correlation rapidly develops 
when the system approaches the valence transition point.
The comparison of the impurity solvers, 
the CT-QMC, non-crossing approximation, and one-crossing approximation, is also addressed.

%
\end{abstract}

\section{Introduction}

Heavy fermion systems with rare-earth or actinide elements 
have attracted considerable attention. 
One of the interesting phenomena in the systems is 
the valence transition, where the valence in rare-earth ions abruptly changes
due to applied field and/or pressure. 
Recently, it has been suggested that 
the superconductivity in the rare-earth compound
$\mathrm{CeCu_2Si_2}$~\cite{Thomas}
and quantum critical behavior in the quasicrystal 
$\mathrm{Au}_{51}\mathrm{Al}_{34}\mathrm{Yb}_{15}$~\cite{Deguchi}
are related to valence fluctuations.
These stimulate further theoretical and experimental investigations
on valence fluctuations in the heavy-fermion systems~\cite{Holmes,Sugibayashi,Watanabe0,Matsukawa,Takemura,Andrade}.

Valence fluctuations and valence transitions have been studied theoretically 
in terms of the extended periodic Anderson model 
(EPAM)~\cite{Onishi,Watanabe1,Saiga,Watanabe2,Kubo,Hagymasi,Kojima}.
The possibility 
 of the valence transition in this model has been suggested by 
the slave boson mean-field theory~\cite{Watanabe2}, density matrix renormalization group~\cite{Watanabe1}, 
and dynamical mean-field theory (DMFT)~\cite{Saiga,Kojima}.
Although the jump singularity in the curve of the $f$-electron number 
has been reported in several papers~\cite{Watanabe1,Watanabe2,Saiga,Kubo,Hagymasi,Kojima}, 
the hysteresis has not been found clearly.
Therefore, it is instructive to examine carefully this problem
by means of unbiased numerical methods.

Motivated by this, we study the EPAM at finite temperatures, 
combining DMFT with a continuous-time quantum Monte Carlo (CT-QMC) 
method~\cite{CTQMC1,CTQMC2,CTQMC3}, 
which is one of the most powerful numerical techniques.
We clarify that the first-order valence transition occurs 
with a hysteresis in the curve of the valence.
We also compare the CT-QMC results with those obtained by 
the non-crossing approximation (NCA)~\cite{Kuramoto1,Kuramoto2,Eckstein} and 
the one-crossing approximation (OCA)~\cite{Eckstein},
and discuss the advantage of the CT-QMC method.

This paper is organized as follows. 
In Sec. \ref{sec:2}, we introduce the model Hamiltonian and 
summarize our numerical method briefly. 
In Sec. \ref{sec:3}, we demonstrate the valence transition 
at finite temperatures in the framework of DMFT with the CT-QMC methods.
We also discuss spin correlation in the vicinity of the valence transition. 
Furthermore, quantitative discussions for the valence transition in the EPAM
are given in Sec. \ref{sec:4}, 
by comparing with the DMFT results obtained by means of the CT-QMC, NCA, 
and OCA impurity solvers.
In the final section, we state our conclusions.

\section{Model and Method}\label{sec:2}
We study valence fluctuations at finite temperatures in the EPAM, which 
should be described by the following Hamiltonian:
\begin{eqnarray}
H_{\mathrm{EPAM}}&=-t\displaystyle\sum_{\langle i,j\rangle,\sigma}
c_{i\sigma}^{\dagger}c_{j\sigma}+V\sum_{i,\sigma}
(c_{i\sigma}^{\dagger}f_{i\sigma}+\mathrm{h.c.})
+\epsilon_f\sum_{i,\sigma}n^f_{i\sigma}  \nonumber \\ 
&+ U_{ff}\displaystyle\sum_{i}n^f_{i\uparrow}n^f_{i\downarrow}
+ U_{cf}\sum_{i,\sigma,\sigma'}n^c_{i\sigma}n^f_{i\sigma'}, 
\end{eqnarray} 
where $c_{i\sigma}$ ($f_{i\sigma}$) is the annihilation operator 
of a conduction electron ($f$-electron) 
with spin $\sigma(=\uparrow,\downarrow)$.
$n^c_{i\sigma}(=c^\dagger_{i\sigma}c_{i\sigma})$ and 
$n^f_{i\sigma}(=f^\dagger_{i\sigma}f_{i\sigma})$ are 
the number operators of the conduction 
and the $f$- electrons at $i$th site, respectively.
$t$ is the hopping integral of the conduction electrons 
between the nearest-neighbor sites,
$V$ is the hybridization between the conduction band and the $f$-orbitals,
and $\epsilon_f$ is the energy level of the $f$-orbitals.
$U_{ff}$ is the repulsive interaction between the $f$-electrons
and $U_{cf}$ is the repulsive interaction between the conduction and the $f$- electrons.
We here consider the Bethe lattice with a large coordination number 
$z\rightarrow\infty$, which has a semi-elliptic density of states 
with the bandwidth $2D$. 

To study low temperature properties in the EPAM~\cite{Yoshida1,Yoshida2}, 
we make use of DMFT~\cite{DMFT1,DMFT2,DMFT3,DMFT4}.
In the framework of DMFT, the lattice model is mapped to an effective impurity 
model, where local electron correlations are taken into account precisely.
This method is formally exact in the infinite dimensions 
and has successfully been used even in three dimensions.
The site-diagonal lattice Green's function is required to be equal to 
the Green's function of the effective impurity model. 
This condition leads to the self-consistent equation~\cite{Schork}, as
\begin{equation}
[{\hat{\mathscr{G}}}^{-1}_{\sigma}(\omega)]_{cc}=\mathrm{i}\omega-\mu-\left(\frac{D}{2}\right)^2[\hat{G}_{\mathrm{imp}\sigma}(\omega)]_{cc},
\end{equation} 
where $\mu$ is the chemical potential, 
${\hat{\mathscr{G}}}_{\sigma}(\omega)$ [$\hat{G}_{\mathrm{imp}\sigma}(\omega)$]
is the noninteracting (full) impurity Green's function with spin $\sigma$. 
To solve the effective impurity model, 
we make use of the CT-QMC method 
based on the hybridization expansion~\cite{CTQMC1,CTQMC2,CTQMC3}. 
In this method, the partition function is described 
by the expansion in powers of the impurity-bath mixing,
which allows us to evaluate physical quantities quantitatively
in terms of the Monte Carlo procedure. 
This is contrast to other biased methods such as the NCA and OCA methods.
In the paper, we set $U_{ff}/D=20$ and $U_{cf}/D=8$,
and use the half bandwidth $D$ as the unit of energy.

\section{First-order valence transition}
\label{sec:3}
Now, we study the EPAM at finite temperatures 
to clarify how valence fluctuations develop 
in the system.
To characterize the Kondo and mixed-valence states, 
we first examine the particle number when the chemical potential is varied.
\begin{figure}[htb]
\centering
\includegraphics[width=38pc]{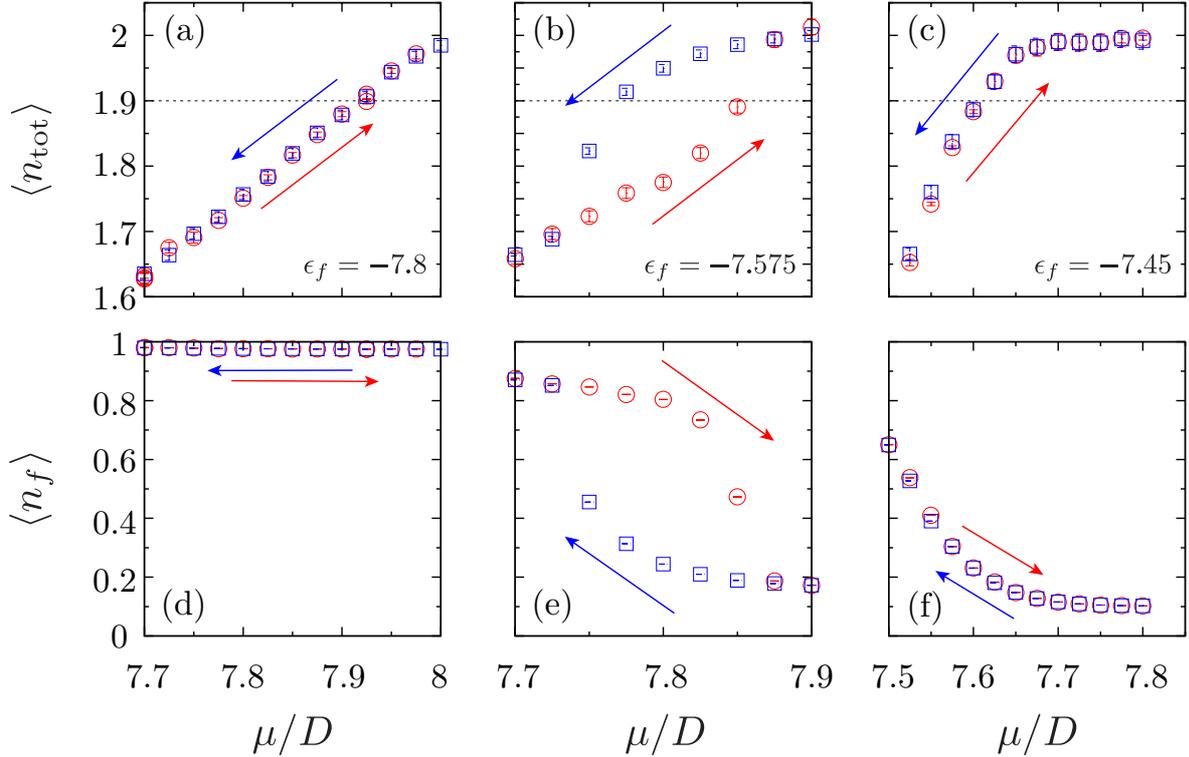}
\caption{\label{label} (Color online)
The total electron number (upper panels) and $f$-electron number (lower panels)
as functions of the chemical potential
in the system with $U_{cf}/D$=$8$, $V/D$=$0.13$, and $T/D=0.04$ 
 when $\epsilon_f/D=-7.8$(left), $-7.575$(middle), and $-7.45$(right).
 The squares (circles) are the data obtained by changing $\mu$ from the higher (lower) side.
}
\label{fig1}
\end{figure}
In Fig.~\ref{fig1}, we show the total particle number per site 
$\langle n_{\mathrm{tot}}\rangle$ and 
the $f$-electron number per site $\langle n_f \rangle$ when $T/D=0.04$ and $V/D=0.13$.
In the case of $\epsilon_f/D=-7.8$,
the total particle number increases monotonically 
with increase of the chemical potential as shown in Fig.~\ref{fig1}(a) 
whereas 
the $f$-electron number little changes within $7.7<\mu/D<8.0$ 
as shown in Fig.~\ref{fig1}(d).
This indicates that the Kondo state is stabilized with localized electrons 
in the $f$-orbitals.
On the other hand, we find in Figs.~\ref{fig1}(c) and~\ref{fig1}(f) that 
both $\langle n_{\mathrm{tot}} \rangle$ and $\langle n_f \rangle$ are smoothly changed
in the case $\epsilon_f/D=-7.45$.
This implies that the mixed-valence state is realized when $7.5<\mu/D<7.8$. 
In the intermediate case with $\epsilon_f=-7.575$, 
remarkably, 
the hysteresis appears 
in the total number of electrons and the $f$-electron number, 
as shown in Figs.~\ref{fig1}(b) and~\ref{fig1}(e).
These results indicate the existence of the first-order valence transition 
in the system
with $\epsilon_f/D=-7.575$.
In fact, if one considers the system with $\langle n_{\mathrm{tot}}\rangle \sim 1.9$,
two solutions are obtained:
the Kondo state with $\langle n_f \rangle \sim 1$ and  
the mixed-valence state with $\langle n_f \rangle \ll 1$.

By performing similar calculations in the system with $\langle n_{\mathrm{tot}} \rangle=1.9$, 
we discuss how the $f$-electron number depends on the energy level of 
the $f$-orbital.
\begin{figure}[htb]
\centering
\hspace*{\stretch{1}}%
\includegraphics[width=27pc]{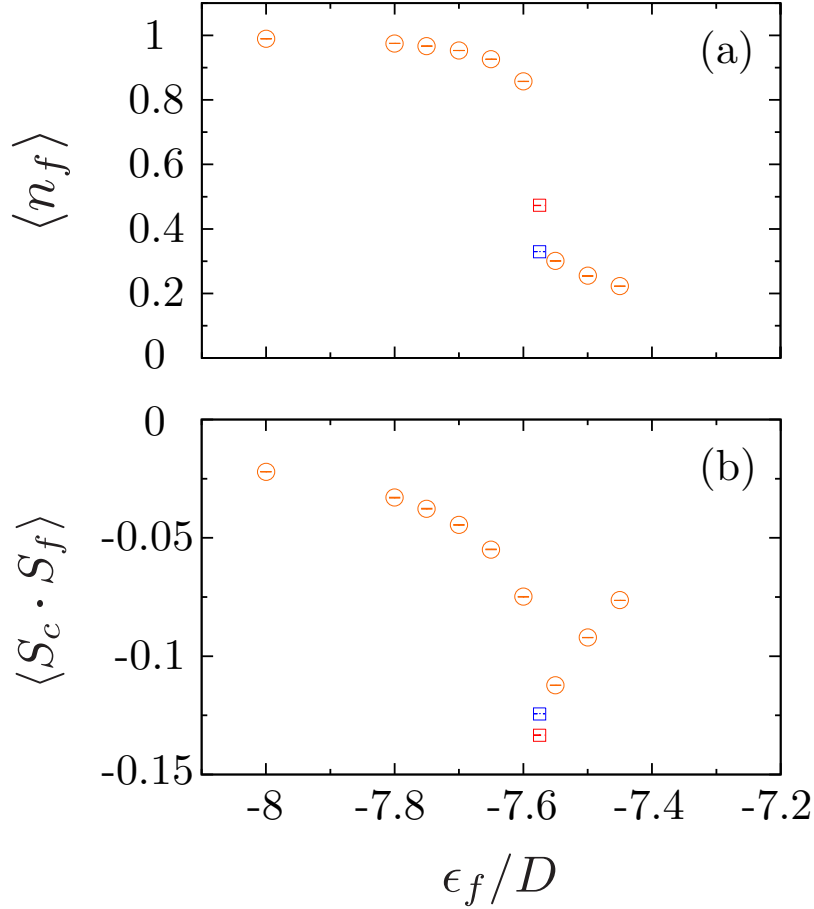}
\hspace*{\stretch{1}}%
\caption{(Color online)
 $f$-level dependence of the $f$-electron 
 number (a) and 
$c-f$ spin correlation (b) when $T/D$=0.04, 
 $U_{cf}/D=8$ and $V/D=0.13$. The squares represent the two solutions at $\epsilon_f/D=-7.575$.
 }
\label{fig2}
\end{figure}
Figure~2(a) shows the $f$-electron number at 
 $T/D=0.04$.
When $\epsilon_f/D=-8.0$, the $f$-electron number is almost one, and
the Kondo state with the localized $f$-electrons is realized.
Increasing $\epsilon_f$, $\langle n_f \rangle$ decreases and 
the first-order valence transition with a jump singularity occurs
at $\epsilon_f\simeq -7.575$.
Beyond the transition point, 
the mixed valence state is stabilized with $\langle n_f\rangle\sim 0.2$.
As discussed above, we have found two solutions 
in the case $\epsilon_f/D= -7.575$. 
Therefore, we can say that
the first-order phase transition with the hysteresis
indeed occurs between the Kondo and mixed-valence states.

\begin{figure}[htb]
\centering
\includegraphics[width=40pc,keepaspectratio]{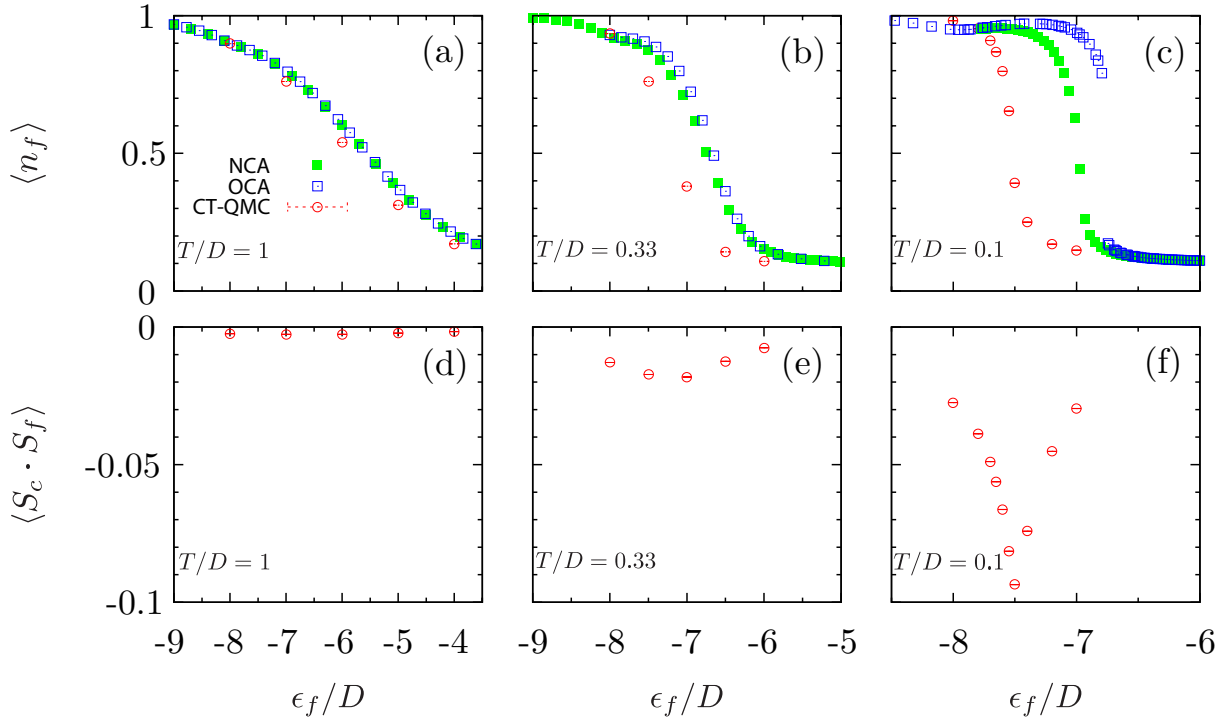}
\caption{(Color online)
 The $f$-electron 
 number (upper panels) and 
$c-f$ spin correlation (lower panels)
in the system with $U_{cf}/D=8.0$ and $V/D=0.15$
when $T/D=1.0$(left), $0.33$(middle), and $0.1$(right). 
The open circles, filled squares, and open squares are the results obtained 
by means of the CT-QMC, NCA, and OCA solvers, respectively.
}
\label{fig3}
\end{figure}

In the vicinity of the valence transition point, 
the singularity appears in other physical quantities. 
Fig.~\ref{fig2}(b) shows the spin correlation 
between the conduction and $f$- electrons $\langle \mathbf{S}_c \cdot \mathbf{S}_f \rangle$.
It is found that the $c-f$ spin correlation is enhanced 
when the system approaches the valence transition point.
Then, the jump singularity appears at the first-order transition point.
In order to clarify how such singularities in the valence and the spin correlation 
develop with decrease of temperature,
we show in Fig.~\ref{fig3} the temperature dependence of the quantities
in the system with $U_{cf}/D=8.0$ and $V/D=0.15$.
When $T/D=1.0$, the valence is smoothly varied and the spin correlation
is little changed with increase of the energy of the $f$-level, as shown in Fig.~\ref{fig3}(d).
This implies that the high temperature paramagnetic state is realized.
The decrease of the temperature leads to the rapid decrease of the valence,
as shown in Figs.~\ref{fig3}(b) and~\ref{fig3}(c).
However, we could not find the jump singularity in the valence 
and thereby a crossover between the Kondo and mixed-valence state occurs. 
In these cases, interesting behavior appears in the spin correlation,
as shown in Figs.~\ref{fig3}(d),~\ref{fig3}(e), and~\ref{fig3}(f).
At 
 $T/D=1.0$, the spin correlation is almost zero and
the Kondo singlet state does not appear.
Decreasing temperatures, the spin correlation is enhanced 
around the crossover region at $T/D=0.33$ and $0.1$, 
as shown in Figs.~\ref{fig3}(e) and~\ref{fig3}(f). 
This indicates that the valence fluctuations enhance the spin correlation 
between the conduction and the localized $f$- electrons.
Further decrease of temperature induces the valence transition
at a certain $\epsilon_f$,
where the cusp singularity should appear in the $c-f$ spin correlation.
Therefore, it is one of the appropriate quantities to discuss 
the nature of the valence transitions.

\section{Comparison with NCA and OCA results}\label{sec:4}

In this section, we compare the CT-QMC method with the NCA and OCA methods.
To this end, we show in the upper panels of Fig.~\ref{fig3}
the $f$-electron number obtained by means of the NCA and OCA solvers. 
At a higher temperature $T/D=1.0$, the system is gradually changed from 
the Kondo like state to the mixed-valence state, as discussed above.
In this case, we could not find a big difference in the curves obtained 
by means of the CT-QMC, NCA, and OCA methods.
Therefore, the NCA and OCA methods are appropriate solvers in the case.
Decreasing temperatures, slightly different behavior appears, 
as shown in Figs.~\ref{fig3} (b) and~\ref{fig3}(c).
We find that the valence crossover points obtained by the NCA and OCA methods
are shifted, comparing with that obtained by the CT-QMC method.
In addition, when $T/D=0.1$, a jump singularity in $\langle n_f \rangle$
appears around $\epsilon_f/D \sim 6.75$, 
which is obtained by means of the OCA method.
This valence transition does not appear 
when the CT-QMC numerical solver is used.
Therefore, we can say that 
the OCA method as well as the NCA method overestimates 
valence fluctuations at low temperatures.

\section{Summary}
We have studied the extended periodic Anderson model, 
combining dynamical mean-field theory with 
the continuous-time quantum Monte Carlo method. 
By calculating the $f$-electron number, 
we have clarified that at low temperatures,
the first-order valence transition 
occurs between the Kondo and the mixed-valence states
together with the hysteresis in the curve of the valence.
We have also found that the $c-f$ spin correlation rapidly develops 
when the system approaches the valence transition point.
In addition, The comparison of the impurity solvers such as 
the CT-QMC, NCA, and OCA has been discussed.

\section{Acknowledgements}
This work was partly supported by the Grant-in-Aid for
Scientific Research from JSPS, KAKENHI No. 25800193.
(A.K.) 
Some of the computations in this work have been
carried out by using the facilities of the Supercomputer Center, 
the Institute for Solid State Physics, the University of Tokyo.
The simulations have been performed by using some of 
ALPS libraries~\cite{alps1}.

\end{document}